\documentstyle[12pt]{article}
\begin{document}
\baselineskip=0.7cm

\begin{flushright} Preprint USM-TH-66 \end{flushright}
\vspace*{.2in}

\begin{center} \Large\bf Number of Electrons per Collision as a Quark-Gluon
Plasma Signal \end{center}
\vspace*{.10in}

\begin{center}
{\underline{C. Dib}}\footnote[1]{e-mail: cdib@fis.utfsm.cl or ischmidt@fis.utfsm.cl},
O. Espinosa\footnotemark[1], P.D. Morley\footnote[2]{Visitor} 
and I. Schmidt\footnotemark[1]  \end{center}
\vspace*{.10in}
\begin{center}
Department of Physics \\ Universidad T\'{e}cnica Federico Santa 
Mar\'\i a \\
Casilla 110-V \\ Valpara\'\i so, Chile  
\end{center}
\vspace*{.10in}
\begin{abstract}
\baselineskip=0.7cm
We investigate whether the total number of electrons per collision,
as a function of beam energy, is a 
potential quark-gluon-plasma (QGP) signature. At high beam energies,
we find that, after experimental removal of the noise induced by Dalitz decays,
the number of produced electrons is increased many times fold 
from the existence of the QGP. The robustness of the
potential signal to differing theoretical assumptions makes it an
attractive experimental parameter.
\end{abstract}

\vspace{.15in}

\newpage
\parskip=9pt

Work is in progress to construct a relativistic heavy-ion collider to
test the theoretical prediction\cite{Uno} that QCD (quantum chromodynamics)
undergoes a phase transition to unconfined quarks and gluons at high
temperatures. Success or failure rides on the identification of an
experimental signature which indicates when a QGP has been achieved.
The complexities of rhic, however, are such that a QCD description from 
first principles will be a long range task. 
Questions of chemical equilibrium, quark effective masses, different flavor 
and gluon temperatures are just some of the many particulars which have 
no universal theoretical consensus. Even a 
parton description is not as adequate for rhic as it is for, say, deeply 
inelastic scattering, since the nuclear medium 
modifies the parton distribution functions and interaction amplitudes. Thus 
from the experimental point of view, a QGP characteristic is desired which is 
independent of the inclusion of the multitude of indeterminate theoretical 
fine minutiae. Original ideas to detect the QGP phase are based on distributions 
of produced leptons and photons\cite{Shu}, and to date much work is being done 
to understand the issues involved\cite{Hung}. However, what we seek here is a robust 
QGP signal, that will survive the fine details. 
We demonstrate that the {\sl total number of direct electrons} 
per collision as a function of beam energy is just such a signature. 

\subsection*{The Result}
In Fig.\ 1 we represent the simulation of the production of electrons in
rhic, after the subtraction of Dalitz decays of the low mass hadrons
$\pi^{\circ}, \; \eta, \; \rho, \; \omega, \; \eta^{\prime}$, $ \phi$,
i.\ e.\ the directly produced electrons.
The Dalitz decays constitute the background (noise) from which the signal
is to be extracted. While the number of mesons expected per collision numbers 
in the thousands (see Fig.\ 2), the accompanying number of directly produced 
electrons, Fig.\ 1, 
can be counted by the fingers of one's hands. Identifying and subtracting 
the Dalitz electron pair background does not represent severe difficulties.

There are two important physical effects associated with the presence of a 
QGP in 
conjunction with the measurement of the total number of directly produced 
electrons per collision, as the beam energy is stepped up. When the energy density 
of the hadron phase first reaches the QGP transition value, allowing a 
mixture of 
the hadronic matter to coexist with the QGP, the number of electrons increases 
more rapidly, with a discernable gradual change in slope around the 
critical beam energy (10-15 GeV/u; see Fig.\ 1): 
this constitutes the first QGP electron signature. Indeed, if one assumes no 
QGP, from a low beam energy region (say, 5 GeV/u) to a high one (say, 30 GeV/u), 
the slope of the lepton yield increases by a factor of 30, while if there were 
a QGP formation at a critical value in between, the slope would increase by a factor
of 200. This is 6 times larger than in the previous case. The fact that there is 
this change in slope is independent of the theoretical uncertainties and is 
due to the increased degrees of freedom associated with the QGP. 
Past the onset of the appearance of the QGP, as shown in Fig.~1, 
the now more rapid electron 
production rate results in a value for the number of directly produced electrons 
per collision many times larger than that in a pure hadron gas: 
this is the second QGP signature. Again, physically, the QGP produces more 
direct electrons 
due to the increased number of degrees of freedom. The simplicity in the 
origin of 
this signal is at the same time its robustness: 
fine details in the model or uncertainties in the parameters 
used cannot cause a misleading signal. Indeed, we have varied the 
values of different parameters of the model over reasonable 
ranges and found no significant change in the two electron signals.
We now discuss how rhic were simulated.

\subsection*{Derivation of the Result}
In central collisions of heavy projectile-target combinations of nuclei, for 
$E_{lab} >$ 1 GeV/u, we expect the formation of thermally 
equilibrated ---or nearly equilibrated---  matter ({\it i. e.} a ``fireball'') 
with subsequent hadronization; for beam energies below 1 GeV/u, no fireball 
should form, since in that case heavy-ion collisions are adequately described 
by the participant spectator model, in which individual nucleons collide while 
the spectator nucleons remain cold and undetected. 
If $E_{lab}$ is increased, the initial temperature of the fireball also 
increases, 
and then subsequently drops down as the fireball progressively expands. 
QCD should undergo a first order phase transition at a given temperature $T_c$.
For gluonic matter, lattice gauge theory suggests a transition temperature 
$T_{c}$ of about 207 MeV. For the temperature ranges involved, it is sufficient 
to consider QCD with just two massless quark flavors (u and d). In describing 
the time evolution of the strongly interacting matter, we take as the equations 
of motion relativistic hydrodynamics\cite{Dos}. 

The picture of a hadronic fireball described by 1-D relativistic
hydrodynamics, with longitudinal expansion and subsequent pion freeze-out, 
has been experimentally verified by
the CERN SPS central S + S collisions at 200 GeV/c per nucleon\cite{Tres}. 
This formulation requires equations 
of state (EOS), the notion of which implies the existence of an equilibrated 
initial uniform piece of nuclear matter.  
Since pionization of nuclear matter occurs already for $T>50$ MeV, the EOS for 
hadronic matter in the relevant temperature range is chosen to be that of a 
(relativistic) pion gas plus a (non-relativistic) nucleon gas. 
The EOS for hadronic matter (pressure $P_{had}$ vs. energy density $\rho_{had}$) 
is given in parametric form by:
\begin{eqnarray} P_{had} =
    \frac{T^4}{6 \pi^2} \int_0^\infty dx 
    \frac {x^4}{\sqrt{x^2+{\tilde m_\pi}^2}}~\sum_{\eta=-1,0,1}~
    \frac{1}{e^{(\sqrt{x^2+{\tilde m_\pi}^2}-{\eta \tilde \mu_\pi})}-1}
                                                                            \nonumber\\  +
    \frac{2T^4}{3 \pi^2} \int_0^\infty dx~ x^4~\sum_{\kappa=p,n}~
    \frac{(2\tilde m_\kappa)^{3/2}}{e^{x^2 +\tilde m_\kappa-\tilde\mu_\kappa}+1}
                                                                             \end{eqnarray}
\begin{eqnarray} \rho_{had} = 
    \frac{T^4}{2 \pi^2} \int_0^\infty dx~ 
    x^2 \sqrt{x^2+{\tilde m_\pi}^2}~\sum_{\eta=-1,0,1}~
    \frac{1}{e^{(\sqrt{x^2+{\tilde m_\pi}^2}-{\eta \tilde \mu_\pi})}-1}
                                                                           \nonumber\\ +
    \frac{T^4}{\pi^2} \int_0^\infty dx~ x^2~\sum_{\kappa=p,n}~
    (\tilde m_\kappa + x^2)\frac{(2\tilde m_\kappa)^{3/2}}
         {e^{x^2 +\tilde m_\kappa-\tilde\mu_\kappa}+1}
                                                                             \end{eqnarray} 
where we denote ${\tilde m_\pi} \equiv m_\pi/T$,  ${\tilde \mu_\pi} \equiv \mu_{\pi^+}/T$, etc.
In each expression, the first sum indicates the contributions from $\pi^-$, $\pi^0$ and 
$\pi^+$, respectively (we use the fact that $\mu_{\pi^+} =      
-\mu_{\pi^-}$, since  $\pi^+$ and $\pi^-$ are mutually antiparticles, 
while $\mu_{\pi^0} = 0$ since $\pi^0$ is its own antiparticle), and the second sum
indicates the contributions from the nucleons ($p=$ proton, $n=$ neutron). 
For numerical purposes, we found it a good approximation for the EOS of the
hadron phase to use $P_{had} = 0.3~\rho_{had}$ in the rhic temperature regime. 

For the quark-gluon plasma, the EOS is determined from a non-inter\-ac\-ting 
relativistic gas of massless $u$ and $d$ quarks and gluons, along with the 
vacuum pressure ($B$ constant), which takes into consideration color 
confinement 
as in a Bag model (this parameter is necessary for the phase transition to be of 
first order). The corresponding EOS is thus described by:
\begin{equation} P_{QGP}={T^{4}\over \pi ^{2}}\int _{0}^{\infty 
}dx\ x^{3}\sum
_{q=u,d}^{\ \ }\left\{ {1\over e^{x-{\tilde \mu} _{q}}+1}+{1\over e^{x+{\tilde \mu}
_{q}}+1}\right\} +{8\pi ^{2}\over 45}T^{4}-B \end{equation}  
\begin{equation} \rho_{QGP}={3 T^{4}\over \pi ^{2}}\int _{0}^{\infty 
}dx\ x^{3}\sum
_{q=u,d}^{\ \ }\left\{ {1\over e^{x-{\tilde \mu} _{q}}+1}+{1\over e^{x+{\tilde \mu}
_{q}}+1}\right\} +{8\pi ^{2}\over 15}T^{4}+B \end{equation}  
where ${\tilde \mu}_q \equiv \mu_q/T$, for $q=u,d$. The first and second terms in the 
sum are the contributions from $q$ and $\bar{q}$, respectively, while 
the last two terms are the contributions from the gluons and the 
(de-confined) vacuum, respectively.
In the case of zero chemical potentials (zero net baryon number and 
electric charge), Eqs. (3) and (4) assume the simple form:
    \begin{equation} P_{QGP}=\frac{37}{3}(\frac{\pi^{2}}{30})T^{4}-B
    \end{equation}
    \begin{equation}
    \rho_{QGP}=37(\frac{\pi^{2}}{30})T^{4} +B \; . \end{equation}
which, for $T_c = 207$ MeV give $B^{1/4} = 288$ MeV. 

In the general case of a net baryon number and/or net electric charge present, 
the equilibrium between the two phases is determined by the three equations in the 
three unknowns $T_{c},\; \mu_{C},\; \mu_{B}$ (critical temperature, electric charge 
chemical potential, baryonic number chemical potential):
    \begin{eqnarray}
    & & P_{had}(\mu_{\pi^{+}},\: \mu_{p},\: \mu_{n},\: T_{c})   \; = \;  
        P_{QGP}(\mu_{u},\: \mu_{d},\: T_{c}) \nonumber \\
     {\cal Q}\; & = & \; {\cal N}_{p} + {\cal N}_{\pi^{+}} -{\cal N}_{\pi^{-}}
     +\frac{2}{3}({\cal N}_{u}-{\cal N}_{\bar{u}})
     -\frac{1}{3}({\cal N}_{d}-{\cal N}_{\bar{d}}) \\
    {\cal B}\; & = & \;  {\cal N}_{p}+{\cal N}_{n}+\frac{1}{3}
    ({\cal N}_{u}-{\cal N}_{\bar{u}})+\frac{1}{3}
    ({\cal N}_{d}-{\cal N}_{\bar{d}})   \nonumber
    \end{eqnarray}
where ${\cal Q}$, ${\cal B}$, ${\cal N}_{p}$, ${\cal N}_{n}$, ${\cal N}_{u}$, 
${\cal N}_{\bar{u}}$, 
\dots, etc are the respective electric charge, baryon number, proton, 
neutron, u-quark, anti-u-quark,\dots, etc number densities. 
The electric charge and baryon number, ${\cal Q}$ and ${\cal B}$, are input 
parameters 
and at the same time the two conserved quantities in the thermal problem, so there are 
only two independent chemical potentials, namely $\mu_C$ and $\mu_B$, in terms of 
which we can express those for each species:
    \begin{eqnarray}
    \mu_{\pi^{+}} \: ( = -\mu_{\pi^{-}}) & =& \mu_{C} \nonumber \\
    \mu_{p} & =& \mu_{C} + \mu_{B} \nonumber \\
    \mu_{n} & =& \mu_{B} \\
    \mu_{u} \; ( = -\mu_{\bar{u}}) & =& \frac{2}{3}\mu_{C}+\frac{1}{3}\mu_{B} \nonumber \\
    \mu_{d} \; ( = -\mu_{\bar{d}}) & =& -\frac{1}{3}\mu_{C}+\frac{1}{3}\mu_{B} \nonumber
    \end{eqnarray}
In the simultaneous solution of these equations, we find that the 
effect of a net baryon number on the value of $T_c$ is negligible, for the 
densities used in the rhic simulation. Thus the important energy densities,
which are shown in Table 1, are fixed by the temperature alone.

As the collision energy is increased, the first appearance of the QGP 
occurs when the energy density reaches $\rho_{\rm had}^{(c)}$ (0.22 GeV/fm$^3$),
at $T_c=207$ MeV. Because the phase transition is modeled to be first 
order, a latent energy density, $\sim 4B$, must then be supplied in order to 
liberate the quarks and reach a 100 \% QGP. This happens when the energy 
density reaches $\rho_{\rm QGP}^{(c)}$ (3.81 GeV/fm$^{3})$. 
If the initial energy 
density, $\rho_{i}$, is between 0.22 and 3.81 GeV/fm$^{3}$, the
matter is in a mixture of the two phases, in which the initial proportion 
of QGP present, $f_{i}$, satisfies 
$\rho_{i}= \rho_{QGP}^{(c)}\times f_{i}+\rho_{had}^{(c)}\times (1-f_{i})$. 

Two parameters describe the collision previous to thermalization: 
the compression factor, $\kappa$, and the stopping power, $S$.
The volume $V_{fb}$ of the initial fireball is taken to be the sum of the 
volumes of the individual nuclei divided by the compression factor:
 \begin{equation} V_{fb}=v_n\times (A_{1}+A_{2})/ \kappa , 
\qquad v_n\simeq 7.2 ~{\rm fm}^{3}/{\rm nucleon}, 
 \end{equation}
where $v_n$ is the nuclear specific volume, and $A_1$ and $A_2$ are the mass 
numbers of the colliding nuclei. The old idea that the fireball can be
infinitesimally small, based on the fact that the colliding nuclei are 
Lorentz contracted `pancakes', is rejected; modern computations disclose 
that a finite time interval is required for equilibration and the compression 
factor $\kappa$ reaches a quasi-constant value of the order of 4.0 at the highest 
energies\cite{Cuatro}. Thus we have adopted a sliding value of $\kappa$ from 1 to 4. 
The other major parameter is the stopping power, $S$, {\it i.e.} the average 
fraction of CMS kinetic energy a nucleon loses in the collision.
For identical nuclei of baryon number A, the CMS thermal energy available for 
hydrodynamic expansion is
\begin{equation}
E_{T}=2A\times(E_N-m_N)\times S ,
\end{equation}
 where $E_N$ is the CMS incident energy per nucleon 
and $m_N$ the nucleon mass.
Experimentally, $S \sim 0.7$ only at high energies, while modest 
for small energies. Thus we also made $S$ a sliding linear function of 
$E_N$. These two parameters are set by the CERN SPS S + S rhic
data\cite{Tres} to be about $S \sim 0.5$ and $\kappa \sim 1.4$ for incident CMS energy per
nucleon $E_{N}$ = 9.7 GeV.

The last consideration is the choice of the relativistic hydrodynamic
description. Event shapes at high energy are not isotropic, but rather
elongated along the beam axis\cite{Tres} so we choose a 1-D longitudinal simulation. 
A phenomenological parton Monte-Carlo simulation\cite{Cinco} gave results
remarkably resembling those of a 1-D longitudinal hydrodynamic flow\cite{Dos}. 

Full numerical solutions including the EOS (1)-(4) and phase coexistence conditions (7)-(8)
were studied. As mentioned above, effects of net baryon charge were found not to be significant, 
in which case the approximate EOS $P_{had} = .3 \rho_{had}$ for hadrons and for Eqs.~(5)-(6) for QGP matter
allow an analytical solution to the 1-D hydro flow: let $\tau = \sqrt{t^2-z^2}$ be the 
proper time of a given fluid element (t,z being the spacetime coordinates from the
origin of collision in the CMS); using $\tau_{0}=$ 1 Fermi/c as the initial time 
({\it i.e.} the estimated time for formation and equilibration of the fireball), for a
high initial temperature the fluid element will start as QGP, cool down as it expands
reaching $T_c$ at $\tau = \tau_1$, condense to a full hadronic phase at $\tau = \tau_2$
and subsequently cool below $T_c$ until freeze out. The temperature $T$ and energy density
$\rho$ in the QGP, mixed and hadronic phases, respectively, are given by:
    \begin{eqnarray} 
        {\rm QGP:}& T(\tau)/T(\tau_{0})=(\tau_{0}/ \tau)^{1/3},~
                            (\rho(\tau)-B)/(\rho(\tau_0)-B) = (\tau_{0}/ \tau)^{4/3}\nonumber\\
        {\rm mixed:}& T(\tau) = T_c ~,~~~\rho_(\tau) = \rho_{QGP}\times f(\tau)
                                                + \rho_{had}\times (1-f(\tau)) \\
        {\rm hadronic:}& T(\tau)/T(\tau_{2})=(\tau_{2}/ \tau)^{0.33},~~~
    \rho_{had}(\tau)/ \rho_{had}(\tau_{2})=(\tau_{2}/ \tau)^{1.3}\nonumber
    \end{eqnarray}
where $f(\tau) = 1 - [1-(\tau_1/\tau)]s_{QGP}/(s_{QGP}-s_{had})$ is the fraction of QGP
in the mixed phase, $s$ being the entropy density $s=(\rho+P)/T$.

The simulation now proceeds as follows: at each beam energy, the initial energy
density and temperature are calculated. Determination is made if the strongly
interacting matter is in the hadronic phase (pions), mixed phase, or pure
QGP phase. An estimate of the prompt electrons is then done, by computing
the di-lepton rate for one time step (units in Fermi/c) over the initial 
fireball volume of a pion gas (the prompt electrons come from the initial fireball 
formation and do not know the existence of any subsequent QGP). After this, the 
hydro-expansion begins: the edges are advanced one time unit using the hydro-solution 
and the fluid is divided into zones. The new temperature is recomputed for the middle 
of each zone and the appropriate di-lepton rate determined, taking into account that
the zone may be in a new phase. The electron rate in the QGP is due to quark-quark  
and quark-gluon collisions\cite{quark-quark} (Fig.\ 3) and in the hadronic phase is 
due to pion annihilation\cite{Seis} and virtual bremsstrahlung from pion 
scattering\cite{Haglin}. Here we did not include heavier 
mesons\cite{resonance} in our analysis, so that we may be slightly 
underestimating the lepton yield from the hypothetical hadronic fireball at high 
temperatures (i.e. above ~$1 GeV$, that is, well above $T_c$). 
Also, finite temperature and density effects were not included, 
since they are small corrections\cite{Siete} to the pion annihilation amplitude in 
vacuum. Those zones which have an energy density less than half-nuclear 
experience freeze-out and do not produce any more particles (electrons and mesons).
The approximate freeze-out temperature is 154 MeV, with the approximate pion
number density .135 fm$^{-3}$.
By integrating over the invariant mass of the lepton pairs and then summing each 
zone contribution, the total production of electrons is ascertained for that 
time interval. We consider a fireball with mirror symmetry around z = 0 and 
cylindrical symmetry about the beam axis. The time counter is advanced an 
additional Fermi/c and the hydro-expansion continues until all zones experience 
freeze-out. The result, Fig.\ 1, for the number of produced electrons as a 
function of the beam energy is obtained. 
As mentioned earlier, the 1-D hydrodynamic model works very well with the
CERN SPS S+S meson data, which constrains the model parameters $\kappa$
and $S$. For those values, Fig.\ 2 shows the pion yield, 
corresponding to the number of pions in the fireball at the moment of freeze-out. 

There is a distinctive increase in the number of directly produced electrons, 
once the beam energy becomes high enough for the QGP phase to form. 
The physical reason for this manifold increase of directly produced 
electrons due to the existence of the QGP is the increased number of degrees of freedom.
This signal is independent of the particulars of the rhic simulation. 
Indeed, for the broad range of values of the phenomenological parameters considered,
namely the stopping power $S$, the compression factor $\kappa$ and the effective gluon
mass, the resulting direct electron signal showed no appreciable change. Thus, one might also 
expect an increased number of hard photons, although the analysis of that signal 
was not considered in this work.

\begin{center} {\bf Acknowledgements}\end{center}
One of us (P.D.M.) is grateful to L. Ray of the STAR detector at RHIC 
for useful discussions. This work was supported in part by FONDECYT (Chile), 
contracts 1950685 and 1960536.

\newpage

\begin{center} {\bf Figure Captions} \end{center}

\begin{enumerate}
\item[1.] The number of electrons per collision versus beam energy (GeV per nucleon),
for Au vs. Au, after subtraction of the Dalitz noise. 
The CMS energy is twice the beam energy, for identical nuclei.
\item[2.] The number of pions per collision versus beam energy (GeV per nucleon), for Au vs. Au.
The CMS energy is twice the beam energy, for identical nuclei.
\item[3.] The processes contributing to the production of dielectrons in the quark-gluon plasma.
\end{enumerate}

\newpage
\begin{center} {\bf Table} \end{center}

\vspace*{0.5in}
\begin{table}[h]
\begin{tabular}{||l|c||} \hline
\mbox{} & \mbox{} \\
$e^{+}e^{-}$ freeze-out ($\sim$ half nuclear energy density) & 0.065 GeV/fm$^{3}$ \\
nuclear energy density & 0.13 GeV/fm$^{3}$ \\
hadron energy density at the phase transition: $\rho_{had}^{(c)}$ & 0.22 
GeV/fm$^{3}$ \\
QGP energy density at the phase transition: $\rho_{QGP}^{(c)}$ & 3.81 GeV/fm$^{3}$   \\
\mbox{} & \mbox{} \\ \hline
\end{tabular}
\caption{Relevant energy densities in rhic.}
\end{table}

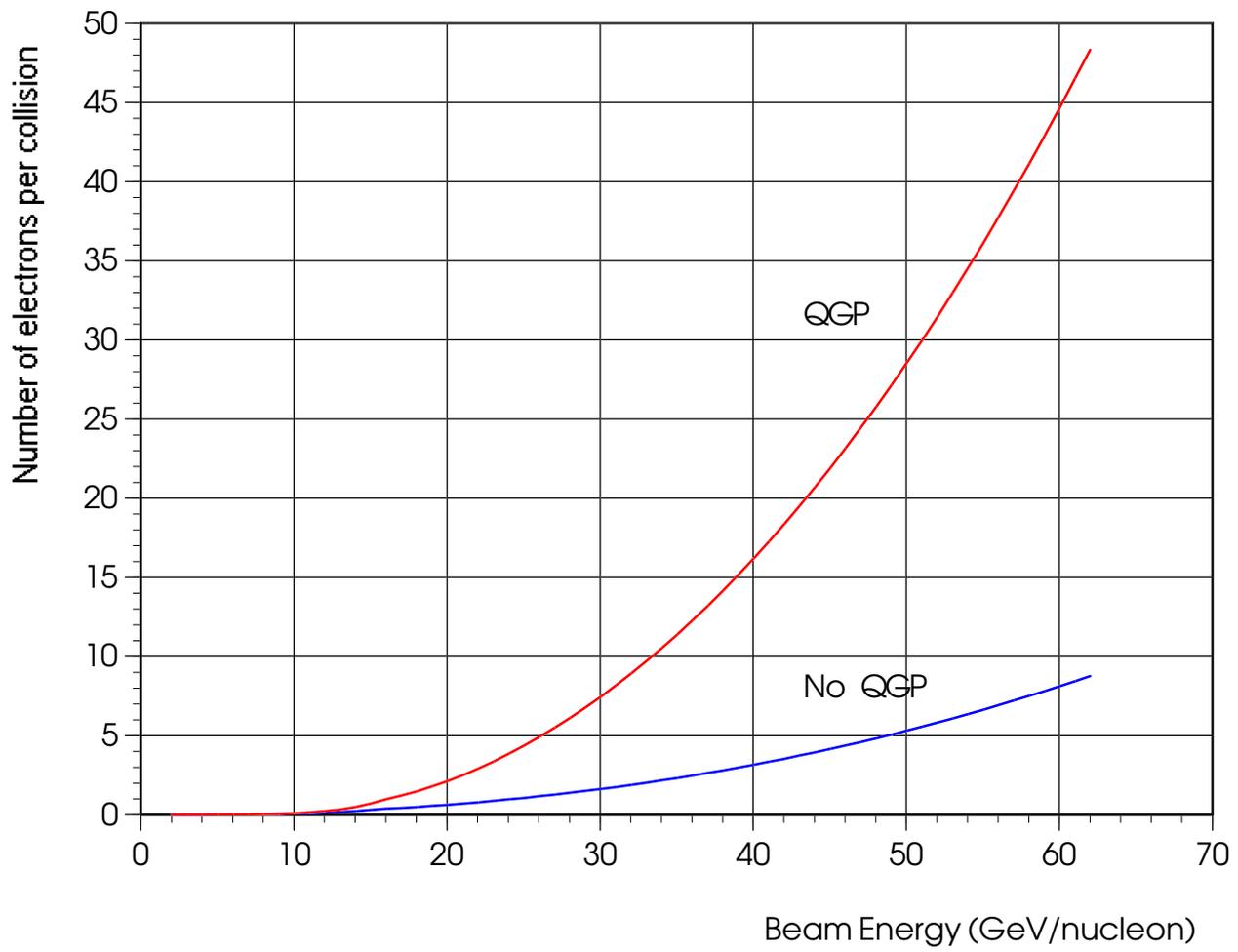
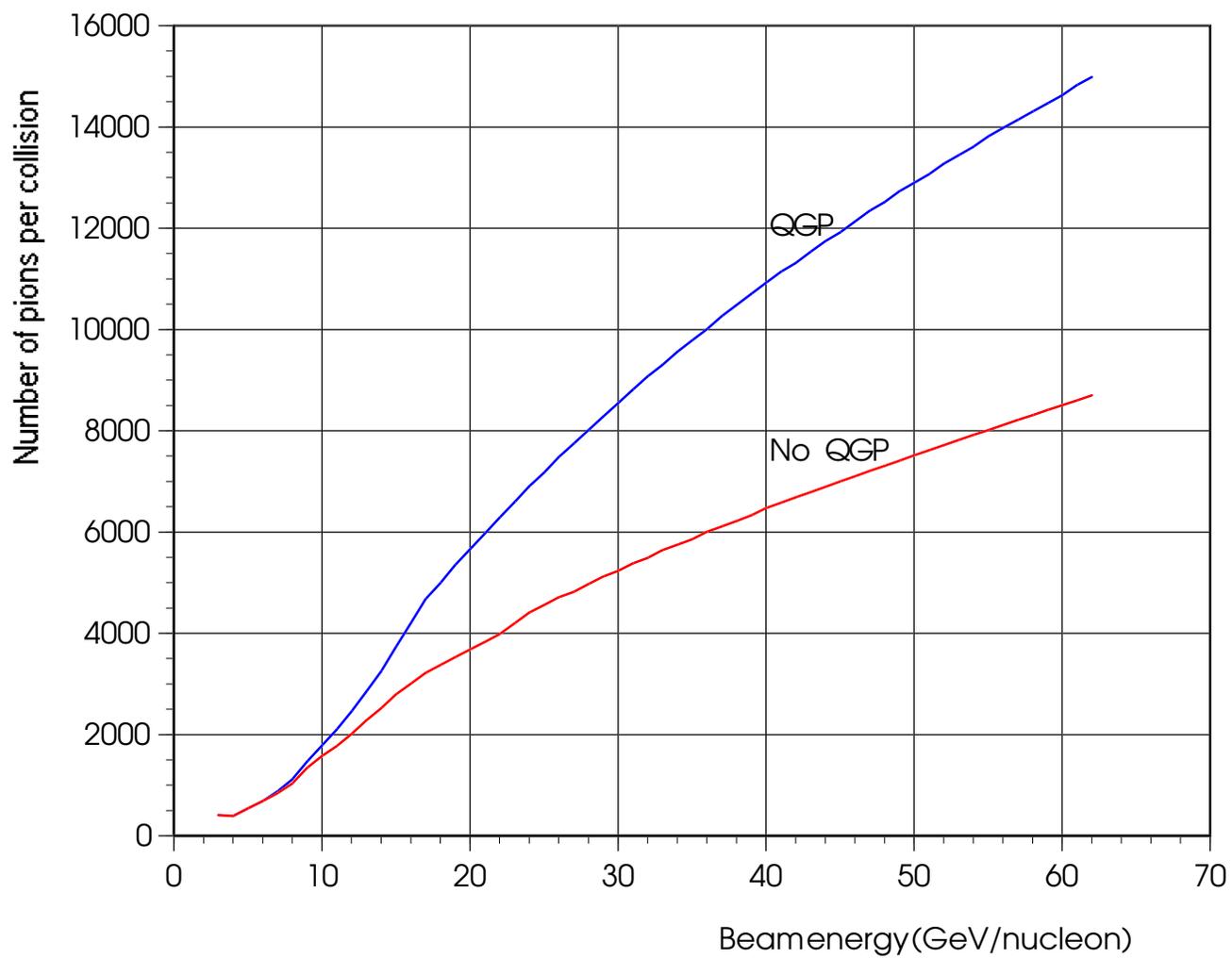
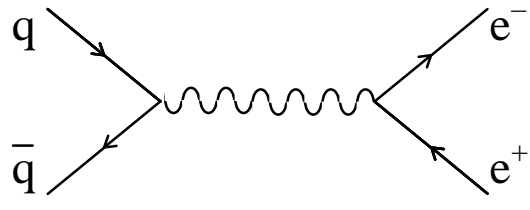
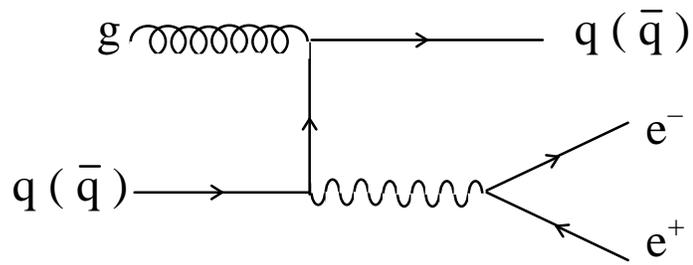
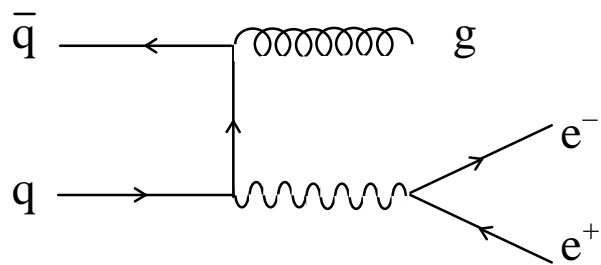
\end{document}